\documentclass[journal,twoside,print]{ieeecolor}
\usepackage{tmi}
\usepackage{cite}
\usepackage{amsmath,amssymb,amsfonts}\usepackage{algorithmic}
\usepackage{graphicx}
\usepackage{textcomp}
\usepackage{xcolor}
\definecolor{revisionorange}{RGB}{0,0,0}
\DeclareRobustCommand{\rev}[1]{{\color{revisionorange}\hypersetup{citecolor=revisionorange,linkcolor=revisionorange}#1}}
\usepackage{multirow}

\usepackage{bm}
\usepackage[english]{babel}
\newtheorem{theorem}{Theorem}

\usepackage{hyperref}
\usepackage{subcaption}
\newcommand{\bqn}{\begin{eqnarray}}
\newcommand{\eqn}{\end{eqnarray}}
\newcommand{\bq}{\begin{eqnarray*}}
\newcommand{\eq}{\end{eqnarray*}}

\newcommand{\ba}{\left( \begin{array}}
\newcommand{\ea}{\end{array} \right)}

\usepackage{hyperref}
\hypersetup{
    colorlinks=true,%
    citecolor=black,%
    filecolor=black,%
    linkcolor=black,%
    urlcolor=black,}

\makeatletter
\renewcommand*\env@matrix[1][c]{\hskip -\arraycolsep
	\let\@ifnextchar\new@ifnextchar
	\array{*\c@MaxMatrixCols #1}}
\makeatother

\def\BibTeX{{\rm B\kern-.05em{\sc i\kern-.025em b}\kern-.08em
    T\kern-.1667em\lower.7ex\hbox{E}\kern-.125emX}}
\begin{document}
\title{\color{black}Minimum Energy Principle \rev{for} \\ \rev{Dynamic Brain Network Interactions}}

\author{Moo K. Chung, Anass B El-Yaagoubi, Anqi Qiu, Hernando Ombao
\thanks{Moo K. Chung is with the University of Wisconsin, Madison, WI 53706 USA., Anqi Qiu is with Hong Kong Polytechnic University, China, Anass B El-Yaagoubi, Hernando Ombao are with KAUST, Saudi Arabia; Correspondence: Moo K. Chung (e-mail: mkchung@wisc.edu).}
}
\maketitle
\pagestyle{plain}
\thispagestyle{plain}

\begin{abstract}
 \rev{ Traditional network models}—such as \rev{granger casuality} and structural equation modeling—\rev{do not explicitly isolate the} cyclic interactions present in dynamic brain networks. We introduce a new \rev{network} framework grounded in the minimum energy principle, where \rev{dissipative} dynamical systems evolve toward configurations of minimal \rev{interaction} energy. In this setting, \rev{network dynamics are modeled as relaxation} from high- to low-energy \rev{configurations}, with interactions represented as flows along connections. Using \rev{network} Hodge theory, we decompose these flows into orthogonal components—gradient and harmonic—each reflecting a distinct topological mode. The gradient \rev{mode dissipates} over time, while the harmonic mode remains as a stable “fixed point,” capturing persistent, non-dissipative cycles in the network. We applied our method to resting-state fMRI brain connectivity data, where it uncovered stable cyclic \rev{interaction} patterns. These results highlight the promise of topological and variational principles for advancing the mathematical foundations of network \rev{interactions}.
\end{abstract}


\section{Introduction}

Understanding \rev{directed interactions} is central to scientific inquiry, as it allows us to \rev{characterize how relationships within a network shape its dynamics} \rev{\cite{zhao.2025}}. In neuroscience, \rev{network} modeling plays a crucial role in uncovering how different regions of the brain \rev{interact with} one another, how information propagates through neural circuits, and how these processes support cognition and behavior \cite{bourakna.2024,lee.2014.MICCAI}. Unlike  correlation-based methods \cite{chung.2019.NN}, which capture symmetric dependencies, \rev{directed interaction} models aim to identify asymmetric relationships that reflect underlying directional \rev{organization} \cite{david.2006,frassle.2021}. This distinction is particularly critical in the analysis of resting-state functional magnetic resonance imaging (rs-fMRI) time series, where temporal \rev{relationships} and recurrent \rev{activity can produce} complex dependencies that challenge conventional modeling assumptions and demand \rev{methods sensitive to cyclic organization}.

Traditional \rev{network} methods such as correlation \cite{mitra.2014}, Granger causality \cite{friston.2014}\rev{, \cite{barnett.2014}, \cite{pelle.2025},} and structural equation modeling \rev{\cite{mcintosh.1994}} have been widely applied to brain imaging data. \rev{Information-theoretic approaches also characterize multivariate and higher-order dependencies \cite{antonacci.2021,antonacci.2025,antonacci.2025.JNE,faes.2022,sparacino.2025,pirovano.2023,faes.2025}. Estimating these dependencies, however, does not by itself isolate the circulation formed collectively by directed interactions. This distinction motivates} the development of the proposed framework \rev{for extracting cyclic structure from estimated directed networks}.

Topological Data Analysis (TDA) has emerged as a promising tool for \rev{characterizing the topology of} complex brain networks. Among its techniques, persistent homology (PH) provides a robust and multiscale framework for capturing \rev{topological} structures by encoding them through simplicial complexes \cite{giusti.2016,lee.2012.TMI}. In this setting, \(0\)-simplices (nodes), \(1\)-simplices (edges), \(2\)-simplices (triangles), and higher-dimensional simplices \rev{encode collections of one, two, three, or more nodes} \rev{\cite{battiston.2020,giusti.2015}. Their simplicial representation does not by itself establish irreducible statistical dependence or synergy.} By tracking the birth and death of such features across multiple scales, PH extracts topological invariants that are both robust to noise and independent of arbitrary resolution choices. This enables a richer characterization of functional brain networks, particularly in resting-state fMRI, than what is afforded by conventional graph-theoretic tools \cite{chung.2019.NN,sizemore.2019}\rev{, \cite{reimann.2017}, \cite{sizemore.2018}}.

\rev{Building on this topological perspective}, we introduce a \rev{network interaction} framework based on the minimum energy principle to detect cyclic interactions in brain networks \rev{\cite{chung.2026.EMBC.causality}}. Our approach \rev{combines an energy-decreasing diffusion with} \emph{network Hodge theory}---a mathematical framework that unifies simplicial topology with spectral geometry \cite{atiyah.1976,lim.2020,anand.2023.TMI,lee.2014.MICCAI}. Central to this framework is the \emph{Hodge decomposition}, which uniquely partitions edge flows into orthogonal components\rev{, gradient and harmonic} \cite{anand.2024.ISBI,lim.2020,su.2024}. This orthogonal separation enables a topology-aware dissection of functional connectivity that \rev{separates source--sink imbalance from balanced circulation. Under the proposed diffusion, gradient interactions dissipate while harmonic circulation remains unchanged. We evaluate cyclic-pattern recovery in randomized networks and assess repeat-scan reproducibility and population-level consistency in rs-fMRI}. Together, these components offer a mathematically rigorous, topologically grounded view of functional brain organization.

\begin{figure}[!t]
\centering
\includegraphics[width=\linewidth]{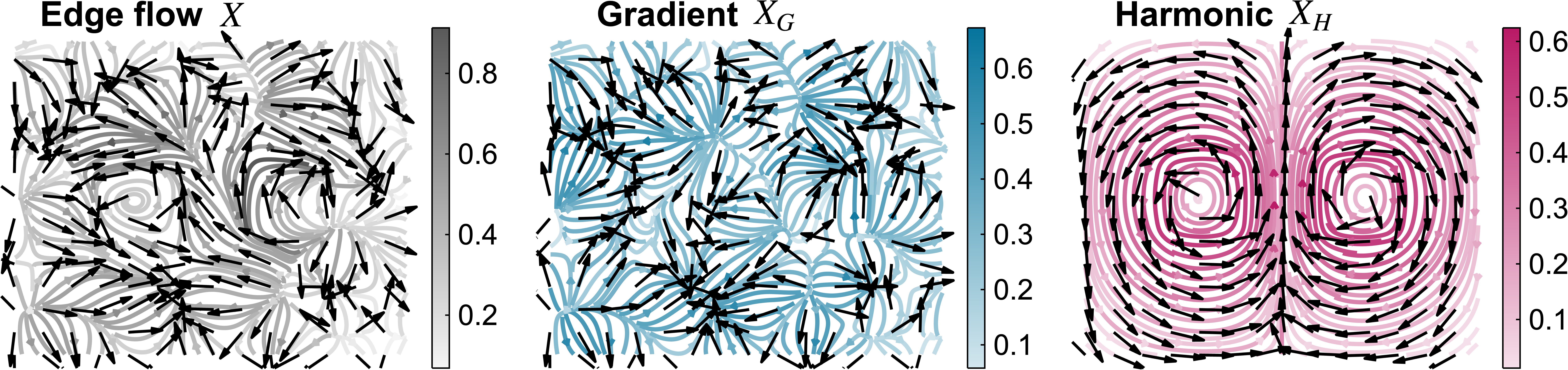}
\caption{\rev{The observed edge flow $X$ contains two counterrotating circular interactions obscured by gradient noise. Hodge decomposition separates the gradient component $X_G$ and recovers the harmonic circulation $X_H$. Streamlines visualize the interpolated fields; colors indicate vector magnitude.}}
\label{fig:grid-gradient-harmonic}
\end{figure}

\section{Methods}

\subsection{Boundary and Coboundary Operators}

In networks, relationships among nodes \rev{and edges are represented by a graph} \cite{anand.2023.TMI}. We assign an orientation to each \rev{edge} using a fixed lexicographic rule: \(i<j\). This orientation is purely a \rev{notational and coding} convention to ensure consistent signs in the construction of boundary and coboundary operators.

The boundary operator \rev{is the node--edge incidence matrix \(\mathbf{B}_1 \in \mathbb{R}^{|V|\times |E|}\). Each column has $-1$ at the tail and $+1$ at the head of its oriented edge, and zero at all other nodes \cite{anand.2024.ISBI}.}
On the other hand, the coboundary operator acts on \rev{scalar functions on nodes and is given by} \( {\bf B}_1^\top \)\rev{, which produces edgewise gradients}. These discrete differential operators form the basis of Hodge theory and support multiscale, geometry-aware analysis of functional signals in brain networks.

The directional network information is then encoded through real-valued functions defined on \rev{edges}. Let \( \mathcal{C}^k \) denote the space of functions on \(k\)-simplices, \rev{with $k=0,1$ for} vertex potentials (\(\mathcal{C}^0\)) \rev{and} edge flows (\(\mathcal{C}^1\)). The boundary operator \(\mathbf{B}_1 : \mathcal{C}^1 \to \mathcal{C}^0\) maps an oriented edge flow to its net incidence at each vertex, i.e., the difference between inflow and outflow. Its kernel,
\[
\ker(\mathbf{B}_1) = \{\, X \in \mathcal{C}^1 : \mathbf{B}_1 X = 0 \,\},
\]
consists of \emph{1-cycles}—closed edge flows with no boundary, where the net inflow equals the net outflow at every vertex.

\subsection{Hodge Decomposition}

The \rev{Hodge decomposition separates edge interactions into gradient and harmonic components (Fig.~\ref{fig:grid-gradient-harmonic}) \cite{anand.2024.ISBI}. The} gradient component exhibits a source–sink structure, and the harmonic component captures \rev{divergence-free cyclic interactions. The decomposition operates directly on estimated directed interactions. These interactions can be obtained from lagged correlation, conditional Granger causality, or DCM.}

\begin{theorem} 
\label{theorem:1}
Let \( X(t) \in \mathcal{C}^1 \) be an edge flow on \rev{a graph}. Then \( X(t) \) admits a unique orthogonal decomposition \cite{anand.2024.ISBI}
\[
\rev{X(t) = X_G(t) + X_H(t),}
\]
where gradient component $X_G(t) = \mathbf{B}_1^\top s(t)$ for some $s(t) \in \mathcal{C}^0$,
\rev{and harmonic component $X_H(t) \in \ker(\mathbf{B}_1)$.}
\end{theorem}

\begin{proof} \rev{Existence}. Given any \( X \in \mathcal{C}^1 \), define the gradient component \( X_G = \mathbf{B}_1^\top s \), where \( s \in \mathcal{C}^0 \) is chosen to solve
\(
\mathbf{B}_1 \mathbf{B}_1^\top s = \mathbf{B}_1 X.
\)
This system is solvable since
\(
\mathbf{B}_1 X \in \operatorname{im}(\mathbf{B}_1) \subset \operatorname{im}(\mathbf{B}_1 \mathbf{B}_1^\top).
\)
Define the harmonic component as the residual
\rev{\(
X_H = X - X_G.
\)}
We show that the residual  \( X_H\) lies in the kernel of \( \mathbf{B}_1 \)\rev{. Note
\(
\mathbf{B}_1 X_H = \mathbf{B}_1 X - \mathbf{B}_1 \mathbf{B}_1^\top s =0.
\)
We used} \( \mathbf{B}_1 \mathbf{B}_1^\top s = \mathbf{B}_1 X \). Hence, the residual satisfies
\rev{\(
X_H \in \ker(\mathbf{B}_1)
\)
and 
\(\mathbf{B}_1X_H = 0.\)
Orthogonality then follows:}
\[
\langle X_H, X_G \rangle = \langle X_H, \mathbf{B}_1^\top s \rangle = \langle \mathbf{B}_1 X_H, s \rangle = 0,
\]
since \( \mathbf{B}_1 X_H(t) = 0 \).

\rev{Uniqueness. Suppose $X$ admits two distinct decompositions $X=X_G+X_H=\widetilde X_G+\widetilde X_H$. Then $X_G-\widetilde X_G=\widetilde X_H-X_H$. Writing $X_G=\mathbf B_1^\top s$ and $\widetilde X_G=\mathbf B_1^\top\widetilde s$, and using that both harmonic components have zero divergence, we obtain}
\[
\rev{\begin{aligned}
\|X_G-\widetilde X_G\|_2^2
&=\langle \mathbf B_1^\top(s-\widetilde s),
\widetilde X_H-X_H\rangle\\
&=\langle s-\widetilde s,
\mathbf B_1(\widetilde X_H-X_H)\rangle=0.
\end{aligned}}
\]
\rev{Thus $X_G=\widetilde X_G$ and consequently $X_H=\widetilde X_H$, contradicting the assumption of distinct decompositions. Hence both components are unique.}
\end{proof}

\rev{The harmonic vector is therefore the unique Euclidean projection of $X$ onto $\ker(\mathbf B_1)$. Since all subjects share the same nodes (brain regions), we use a common complete graph with fixed reference edge orientations. Thus, the incidence matrix $\mathbf B_1$ and the harmonic projection operator remain fixed across subjects, models, and window settings}.

\rev{Although Hodge decomposition extends to higher-dimensional simplicial complexes \cite{lim.2020}, our focus is modeling directed cyclic interactions in brain networks. A directed cycle is a closed sequence of directed edges and is therefore a one-dimensional object. We accordingly use the graph-based Hodge decomposition of edge flows in $\mathcal{C}^1$.}

\subsection{\rev{Numerical Estimation}}
\rev{The proof of Theorem~\ref{theorem:1} gives $X_G=\mathbf B_1^\top s$ with $\mathbf B_1\mathbf B_1^\top s=\mathbf B_1X$. The pseudoinverse acts as an inverse on the range of the singular matrix $\mathbf B_1\mathbf B_1^\top$, which contains $\mathbf B_1X$. Thus $s=(\mathbf B_1\mathbf B_1^\top)^\dagger\mathbf B_1X$ solves the system, giving $X_G=P_GX$, with}
\[
\rev{P_G = \mathbf{B}_1^\top (\mathbf{B}_1 \mathbf{B}_1^\top)^\dagger \mathbf{B}_1,}
\]
where \( (\cdot)^\dagger \) denotes the Moore–Penrose pseudoinverse. The harmonic \rev{residual satisfies $X_H=X-X_G=(I-P_G)X$. Hence, the harmonic projection operator is $P_H=I-P_G$:}
\[
\rev{P_H = I - \mathbf{B}_1^\top (\mathbf{B}_1 \mathbf{B}_1^\top)^\dagger \mathbf{B}_1.}
\]
\rev{The harmonic} component is then computed by
\rev{$
X_H = P_H X
$} {\rev{\cite{chung.2026.ISBI}.}}
Note \rev{$P_G$} and \rev{$P_H$ satisfies}
$$\rev{P_G + P_H = I, \quad P_G P_H = P_H P_G = 0.}$$
\rev{For the complete graph used in the rs-fMRI analysis, these projection formulas simplify as follows.}

\begin{theorem}
\label{theorem:complete-graph-lse}
\rev{Let $\mathbf B_1\in\mathbb R^{|V|\times|E|}$ be the boundary matrix of a complete graph on $|V|$ nodes. For edge vector $X\in\mathbb R^{|E|}$, the gradient component $X_G=P_GX$ and harmonic component $X_H=P_HX$ are given by}
\[
\rev{\begin{aligned}
X_G&=\frac{1}{|V|}\mathbf B_1^\top\mathbf B_1X, \quad X_H=X-\frac{1}{|V|}\mathbf B_1^\top\mathbf B_1X.
\end{aligned}}
\]
\end{theorem}

\begin{proof}
\rev{Each column of $\mathbf B_1$ contains one $+1$ and one $-1$,} so \rev{$\mathbf 1^\top\mathbf B_1=0$. Each node belongs to $|V|-1$ edges, and each pair of distinct nodes shares one edge with opposite incidence signs. Thus, \(\mathbf{B}_1\mathbf{B}_1^\top=|V|P\), where \(P=I-\mathbf{1}\mathbf{1}^\top/|V|\) is an orthogonal projector. Since \(P^\dagger=P\),}
\[
\rev{(\mathbf{B}_1\mathbf{B}_1^\top)^\dagger
=\frac{P}{|V|}
=\frac{1}{|V|}\left(I-\frac{\mathbf{1}\mathbf{1}^\top}{|V|}\right).}
\]
\rev{Substituting into \(X_G=P_GX\) and using \(\mathbf{1}^\top\mathbf{B}_1=0\) gives}
\[
\rev{X_G=\frac{1}{|V|}\mathbf{B}_1^\top P\mathbf{B}_1X
=\frac{1}{|V|}\mathbf{B}_1^\top\mathbf{B}_1X.}
\]
\rev{The harmonic component follows from $X_H=X-X_G$.} 
\end{proof}

\rev{The harmonic projection requires only sparse incidence operations, taking $O(|E|)$ time and $O(|E|+|V|)$ working storage. Mean harmonic-projection times on edge vectors were 0.027, 0.332, and 2.032 ms for complete graphs with 116, 400, and 1000 nodes, respectively, using MATLAB R2026b on macOS ARM64. The 400- and 1000-node benchmarks match the resolutions of the Schaefer-400 and Schaefer-1000 cortical parcellations \cite{schaefer.2018}. Timings were averaged over 400 sequential batches of 100 projections. Each 1200-frame scan requires 1145 harmonic projections using 56-frame windows with one-TR sliding. Across all 1543 scans, estimated total projection times are 47.4 seconds, 9.8 minutes, and 59.8 minutes for 116-, 400-, and 1000-region parcellations, respectively.}

\subsection{\rev{Minimum Energy Principle for Directed Interactions}}

\rev{The minimum-energy principle provides a general mathematical model of dissipative dynamics in interacting systems \cite{chung.2026.ISBI}, with neuronal interactions as one application \cite{chung.2026.EMBC.causality}. The Taylor expansion of system's energy $E$ near a stable equilibrium $x_0$ gives}
\[
\rev{E(x)-E(x_0)
\approx \frac{1}{2}(x-x_0)^\top H(x_0)(x-x_0),}
\]
\rev{where $H(x_0)$ is the Hessian at equilibrium. For pairwise interactions modeled by quadratic differences between node deviations from equilibrium, the Hessian takes the form of a weighted node Laplacian $\mathcal{L}_0=D-W$, with symmetric nonnegative weights $W=(w_{ij})$ and $D_{ii}=\sum_j w_{ij}$. The quadratic approximation is then a Dirichlet energy. Energy-based descriptions of neural dynamics also have a precedent in Hopfield networks \cite{hopfield.1984}. Graph diffusion has also been used to relate anatomical connectivity to resting-state functional connectivity \cite{abdelnour.2014}.}

An edge flow \(X \in \mathcal{C}^1\) can be interpreted as a fluid-like transport or information flow across the network. \rev{Here, the energy principle defines a mathematical diffusion of directed interactions; it does not establish predictive or interventional causality. For interactions represented on network edges, the corresponding quadratic energy operator is the \emph{1-Hodge Laplacian} $
\mathcal{L}_1 = \mathbf{B}_1^\top \mathbf{B}_1
$}
and the associated Dirichlet energy
\[
\rev{E(X) = \frac{1}{2} \langle X, \mathcal{L}_1 X \rangle = \frac{1}{2} \| \mathbf{B}_1 X \|_2^2.}
\]
The term \( \| \mathbf{B}_1 X \|_2^2 \) measures divergence, capturing net inflow or outflow at nodes (i.e., source-sink behavior)\rev{. As an optimization principle, this resembles variational free-energy theory; however, variational free energy is defined through a probabilistic generative model \cite{friston.2010}, whereas $E(X)$ measures interaction imbalance. The diffusion evolves} flows toward states that minimize \rev{this} energy. Just as water flows downhill from high potential energy to low, \rev{the diffusion} is driven by the gradient of the latent energy:
\[
\nabla E(X) =  \nabla \frac{1}{2} X^{\top} \mathcal{L}_1 X  = \mathcal{L}_1 X.
\]
Thus, gradient descent on \( E(X) \) leads to the diffusion equation
\begin{equation}
\rev{\frac{dX(t)}{dt} = -\nabla E(X(t)) = -\mathcal{L}_1 X(t),
\label{eq:diffusion}}
\end{equation}
which evolves \( X(t) \) in the direction of steepest energy descent \rev{\cite{chung.2026.EMBC.causality}. In the brain-network application, the diffusion is initialized with the full edge-weight vector $X(0)$, without thresholding.} Along this trajectory, the Dirichlet energy decreases monotonically:
\[
\frac{d}{dt} E(X(t)) = \left\langle \frac{dX}{dt}, \mathcal{L}_1 X(t) \right\rangle = - \| \mathcal{L}_1 X(t) \|^2 \leq 0.
\]

Under this dynamics along the \rev{trajectory}, each Hodge component evolves independently:
\begin{align*}
\rev{\frac{d X_G(t)}{dt}} &\rev{= -\mathbf{B}_1^\top \mathbf{B}_1 X_G(t).}
\end{align*}
However, the harmonic component is unaffected by the dynamics:
\rev{$
\frac{d X_H(t)}{dt} = 0.
$}
The harmonic component lies entirely in the null space of the Hodge Laplacian and is therefore unaffected by the diffusion flow \rev{\cite{chung.2026.ISBI}}. It encodes persistent, non-dissipative cycles in the flow. These cycles reflect globally constrained, recurrent structures that remain invariant under energy (or information) dissipation, revealing stable topological patterns that persist throughout the evolution.

Solving each equation yields the time-evolved components:
\[
\rev{X_G(t) = e^{-\mathbf{B}_1^\top \mathbf{B}_1 t} X_G(0),}
\]
and $X_H(t) = X_H(0)$.
\rev{The} gradient component \( X_G(t) \) \rev{decays} exponentially to zero. In contrast, the harmonic component \( X_H(t) \) remains invariant. \rev{The limiting edge flow is therefore $X_H=\lim_{t\to\infty}X(t)=P_HX(0)$,} revealing stable, large-scale, and recurrent structures in the network. These harmonic flows serve as the {\it fixed points} of the dynamics, representing topologically conserved modes that resist decay and thus encode persistent \rev{cyclic interactions}.

\rev{In rs-fMRI, this formulation investigates how spontaneous temporal relationships organize into circulation across brain regions. The proposed diffusion models the relaxation of estimated interactions, reducing the imbalance between incoming and outgoing interactions while preserving balanced cycles. Ongoing endogenous neural input can change this organization across windows \cite{chung.2026.EMBC.causality}. Here, Dirichlet energy quantifies interaction imbalance rather than the metabolic energy consumed by the brain.}

\rev{Figure~\ref{fig:grid-gradient-harmonic} illustrates how the proposed diffusion removes interaction imbalance while preserving circulation. On a grid with 725 nodes and 1374 edges, we added gradient interactions to two counterrotating circular patterns, producing the observed edge flow $X$. The two patterns mimic left- and right-hemisphere circulation in rs-fMRI we obtained in the study. Harmonic projection separates the source--sink imbalance $X_G$ from the underlying cyclic pattern $X_H$, recovering both circulations. Under the proposed diffusion, the gradient component decays while the harmonic component remains unchanged.}

\subsection{\rev{Stability of Harmonic Projection}}
\label{sec:harmonic-stability}

\rev{We quantify the stability of harmonic flow under perturbations. The harmonic component $X_H=P_HX$ is the orthogonal projection of $X$ onto $\ker(\mathbf B_1)$ \cite{lim.2020}. For a perturbation $\delta X$, the perturbed harmonic component $\widetilde X_H=P_H(X+\delta X)$ satisfies}
\[
\rev{\|\widetilde X_H-X_H\|_2
=\|P_H\delta X\|_2
\leq\|\delta X\|_2.}
\]
\rev{Thus, harmonic projection cannot amplify perturbations \cite{parikh.2014}. Since the gradient and harmonic subspaces are orthogonal, $P_HP_G=0$. Any gradient perturbation $\delta X_G=P_G\delta X$ is therefore removed entirely:}
\[
\rev{P_H(X+\delta X_G)=P_HX=X_H.}
\]
\rev{This filtering property provides robustness to gradient perturbations, while harmonic perturbations are preserved.}

\subsection{\rev{Spectral Permutation Test for  Cyclic Interactions}}
\label{sec:population-spectral-test}

\rev{We developed a spectral permutation test \cite{wang.2022} for a shared population-level signed cyclic pattern. Individual cyclic patterns can be nonzero yet cancel across subjects. We write the common harmonic projector as}
\[
\rev{P_H=UU^\top,\qquad U^\top U=I,\qquad \mathbf B_1U=0.}
\]
\rev{The columns $u_k$ of $U=[u_1,\ldots,u_q]$ form a fixed orthonormal cycle basis. Let $X_{H,i}$ be subject $i$'s harmonic vector averaged over windows within each scan and then over that subject's available scans. We write $X_{H,i}=\sum_{k=1}^{q}a_{ik}u_k$, where $a_{ik}=u_k^\top X_{H,i}$ is the scalar coefficient of component $k$. We estimate the population cyclic pattern $\mu_H=\mathbb E[X_{H,i}]$ by $\bar X_H=N^{-1}\sum_{i=1}^{N}X_{H,i}$ and test}
\[
\rev{H_0:\mu_H=0\quad\text{vs.}\quad H_1:\mu_H\ne0.}
\]
\rev{Under $H_0$, there is no common signed population pattern. We treat subjects as independent and assume no preferred overall circulation sign: an individual pattern and its sign reversal are equally likely, $X_{H,i}\overset{d}{=}-X_{H,i}$ \cite{winkler.2014}. We use the squared magnitude of the sample mean. By orthonormality of the cycle basis,}
\[
\rev{T=\|\bar X_H\|_2^2
=\sum_{k=1}^{q}\left(\frac{1}{N}\sum_{i=1}^{N}a_{ik}\right)^2.}
\]
\rev{Larger values indicate that subjects' signed cyclic patterns reinforce one another rather than cancel. The statistic is invariant to the choice of orthonormal cycle basis. Independence and} the \rev{sign symmetry imply}
\[
\rev{(X_{H,1},\ldots,X_{H,N})
\overset{d}{=}(\epsilon_1X_{H,1},\ldots,\epsilon_NX_{H,N})}
\]
\rev{for any fixed signs $\epsilon_i\in\{-1,1\}$ \cite{hemerik.2018}. We draw the signs independently with equal probability, using one sign for all components of each subject:}
\[
\rev{\begin{gathered}
X_{H,i}^{*}=\epsilon_iX_{H,i},\quad
\bar X_H^{*}=\frac{1}{N}\sum_{i=1}^{N}X_{H,i}^{*}, \quad
T^{*}=\|\bar X_H^{*}\|_2^2.
\end{gathered}}
\]
\rev{Each randomized vector remains harmonic and retains its norm, while its contribution to the population mean can reverse}.

\rev{The statistic admits online updates \cite{chung.2019.transposition}. Let $S_k=\sum_{i=1}^{N}\epsilon_i a_{ik}$, so $T^{*}=N^{-2}\sum_{k=1}^{q}S_k^2$. Reversing subject $i$'s sign changes every component sum, giving}
\[
\rev{\begin{aligned}
S_k'=S_k-2\epsilon_i a_{ik},\quad
T^{*\prime}=T^{*}+\frac{1}{N^2}\sum_{k=1}^{q}\big[(S_k')^2-S_k^2\big],
\end{aligned}}
\]
\rev{where primes denote updated values and $\epsilon_i$ is the sign before reversal. All components are updated before evaluating the new sign configuration. For $B$ independent sign configurations, the $p$-value is
$p=(1+\#\{b:T^{*(b)}\ge T\})/(B+1)$.}
\section{\rev{Validation}}

\label{sec:simulation-comparison}

\subsection{\rev{Ground Truth Cyclic Interactions}}
\begin{figure}[t]
\centering
\includegraphics[width=\linewidth]{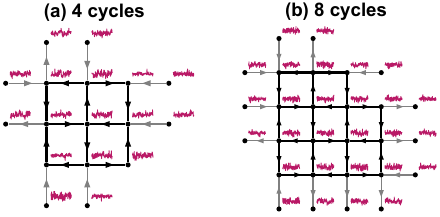}
\caption{\rev{Four- and eight-cycle examples} with \rev{17} and \rev{28 nodes}. \rev{Thicker black edges form the ground-truth cyclic interactions; gray edges are noncyclic interactions}. \rev{Pink traces show} the \rev{first 100 simulated time points}.}
\label{fig:simulation-circulation}
\end{figure}

\rev{We generated 10,000 randomized square-grid networks at each noise level, containing 6--115 nodes and using 1,000 realizations for each independent cycle count from 1 to 10. Each network contained the specified number of randomly arranged square cycles, with neighboring cycles circulating in opposite directions. Tree branches were then added using one of three growth rules, each selected with probability $1/3$, with uniform selection among eligible connections. Each branch direction was chosen with equal probability. Figure~\ref{fig:simulation-circulation} shows two representative examples.}
 
Time series data were then simulated such that their causal dependencies followed the ground-truth flow by generating a vector autoregressive (VAR) 
process of order $L=10$ \rev{\cite{papana.2017} \cite{primiceri.2005}}:
\[
Z_t \;=\; \sum_{\ell=1}^L A_\ell Z_{t-\ell} + \varepsilon_t, 
\qquad \varepsilon_t \sim \mathcal{N}(0, \sigma^2 I)
\]
with $200$ time points per simulation \rev{and $\sigma=1$ or $10$}. Longer lags ($L=10$) progressively obscure the direct cycle structure, making the causal dependencies more difficult to recover. The directed flow matrix 
$W = (w_{ij})$ was constructed by matching the edge flows ($i \to j$). Each lag matrix $A_{\ell}$ was given by
\rev{\(
A_1=\rho I+cW^\top
\)
and
\(
A_{\ell}=c\cdot\delta^{\,\ell-1}W^\top
\)
for $\ell=2,\ldots,L$,}
where $\rho=0.1$ controls the self-retention strength (the extent to which each node preserves its own past activity), 
$c=0.1$ sets the interaction strength along the directed edges (scaling how strongly past activity of a source node drives its neighbors), 
and $\delta=0.5$ imposes geometric decay in the contribution of longer lags. This formulation embeds the ground-truth edge flow into the temporal dynamics. 

\subsection{\rev{Estimation and Recovery of Directed Cycles}}
\label{sec:estimation-recovery}
\rev{We compare lagged correlation \cite{mitra.2014}, conditional Granger causality \cite{barnett.2014}, and SEM \cite{mcintosh.1994}, each before and after harmonic projection. For lagged correlation, $C_{i\to j}=\operatorname{corr}(Z_i(t),Z_j(t+1))$ is the lag-one Pearson correlation. We use the forward--backward difference rather than selecting a dominant direction:}
\begin{equation}
\rev{\label{eq:lagged-correlation-contrast}
X_{ij}=\frac{C_{i\to j}-C_{j\to i}}{2}.}
\end{equation}
\rev{Positive and negative contrasts indicate opposite net directions; equal correlations give zero. The contrast varies continuously with both estimates and cancels contributions common to the two directions. This avoids the abrupt switch that occurs when nearly equal estimates exchange dominance.

For conditional Granger causality \cite{barnett.2014,seth.2015}, we compute the directed edge weight $G_{i\to j}$ as the logarithm of the reduced-to-full prediction-error variance ratio. The full model includes the past $p$ observations of every node; the reduced model removes only the source's $p$ lagged predictors. We selected $p$ among estimable orders up to 10 by minimizing prediction error using a chronological 70\% training and 30\% validation split. SEM retains its signed path coefficients. All methods use the same recording and supplied edges, without ground-truth cycle labels.

As a performance metric, we used cosine similarity \cite{chung.2026.EMBC.causality,luo.2018}, $\widehat X^\top X^\star/(\|\widehat X\|_2\|X^\star\|_2)$, where $\widehat X$ is the estimate before or after harmonic projection and $X^\star$ is the known cyclic ground truth. Uniform positive rescaling preserves relative edge weights and interaction directions. Cosine similarity therefore compares cyclic patterns independently of their overall magnitude, measuring agreement with the known circulation across all supplied edges}.

\begin{figure}[!t]
\centering
\includegraphics[width=\linewidth]{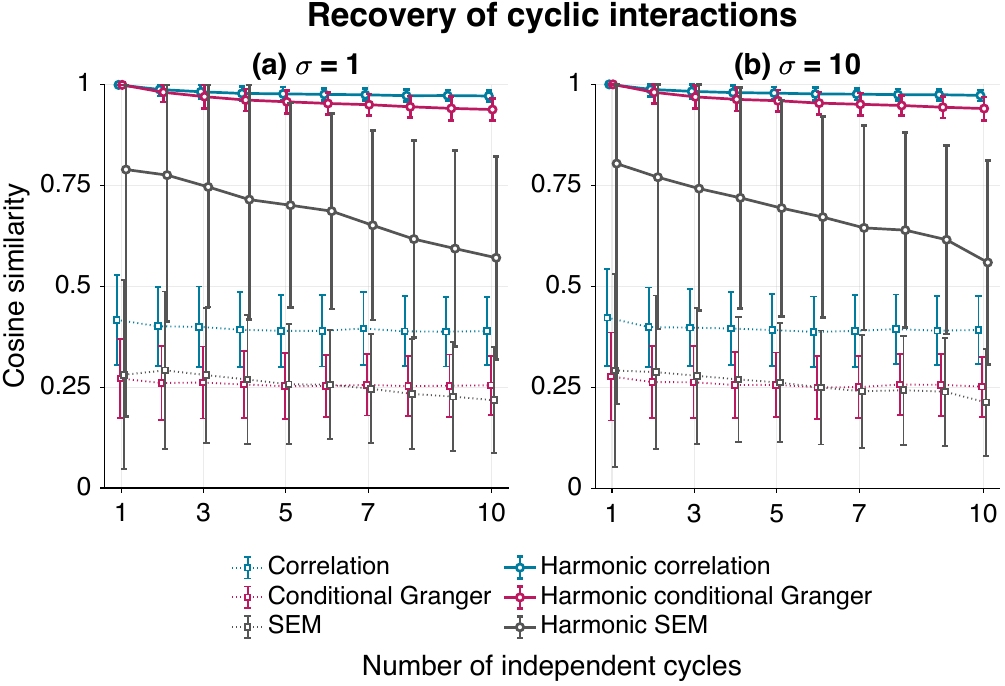}
\caption{{\bf \rev{Recovery of cyclic interactions}}\rev{: mean cosine} similarity \rev{and sample SD across 1,000 random networks per cycle count for} (\rev{a}) \rev{$\sigma=1$} and (\rev{b}) \rev{$\sigma=10$}. \rev{Each panel uses independently generated networks, weights,} and \rev{Gaussian noise. Dotted curves show} the \rev{original estimates; solid curves show their harmonic projections}.}
\label{fig:simulation-cyclic-recovery}
\end{figure}

\rev{Figure~\ref{fig:simulation-cyclic-recovery} shows the results. At ten cycles and $\sigma=1$, harmonic projection increases mean cosine similarity from 0.389 to 0.973 for correlation (150.0\% improvement), from 0.255 to 0.939 for conditional Granger (268.5\% improvement), and from 0.218 to 0.571 for SEM (162.4\% improvement). These substantial gains reflect the ability of harmonic projection to eliminate branch contributions while preserving cyclic interactions. Independent simulations at $\sigma=10$ gave similar results; at ten cycles, mean cosine similarities after harmonic projection were 0.973 for correlation, 0.940 for conditional Granger, and 0.559 for SEM}.

\section{Application}
\subsection{Functional Brain Imaging Data and Preprocessing}

We analyzed resting-state functional magnetic resonance imaging (rs-fMRI) data from 400 healthy adults (168 males, 232 females; age range 22–36 years, mean $29.24 \pm 3.39$ years) obtained from the Human Connectome Project (HCP) \cite{vanessen.2012}. Data were acquired on Siemens 3T Connectome Skyra scanners using a gradient-echo EPI sequence (multiband factor $8$, repetition time (TR) $=720$ ms, echo time (TE) $=33.1$ ms, flip angle $52^\circ$, matrix size $104 \times 90$ (RO $\times$ PE), $72$ slices, $2$ mm isotropic voxels). Each run lasted approximately $15$ minutes and contained $1200$ time points. Participants were instructed to rest with eyes open, fixating on a cross-hair projected on a dark background. Imaging data processed through the HCP's minimal preprocessing pipeline \cite{glasser.2016}, which included spatial normalization and motion correction. Volumes with framewise displacement greater than $0.5$ mm, along with their immediate neighbors, were scrubbed \cite{power.2012}; subjects with excessive head motion were excluded. The Automated Anatomical Labeling (AAL) atlas \cite{tzourio.2002} was used to parcellate brain into 116 non-overlapping anatomical regions. For each subject, the rs-fMRI signals were averaged across all voxels within each region, producing 116 regional time series of length $1200$ \cite{huang.2020.NM,song.2023}.

\rev{We represent these regional rs-fMRI signals as} a multivariate time series \rev{\(X(t) = [X_1(t),\ldots,X_{|V|}(t)]\)}, where each \(X_i(t)\) represents the activity of brain region \rev{\(i\). Directed interactions are estimated using lagged correlation \cite{mitra.2014,chung.2026.EMBC.causality}, We use the forward--backward difference in} {\rev{\eqref{eq:lagged-correlation-contrast}}} \rev{rather than selecting a dominant direction. As a mathematical convention, we use the lexicographic rule to assign a reference orientation $i\to j$ for $i<j$, defining the signs in $\mathbf B_1$ and the edge coordinates. This convention does not impose an estimated interaction direction or establish causality.}

\begin{figure}[t!]
\centering
\includegraphics[width=\linewidth]{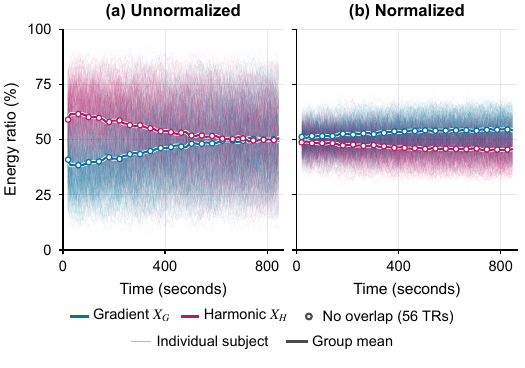}
\caption{Temporal evolution of Hodge decomposition in dynamic resting-state fMRI brain networks. \rev{Component vectors were averaged across repeat scans within each subject: (a) without normalization and (b) after normalizing each component within each scan and window. Thin curves} represent individual trajectories, while thick \rev{curves} denote the mean ratios across subjects.}
\label{Fig:loopanalysis}
\end{figure}

\subsection{Emergence of the Stable Cyclic Structures in rs-fMRI}

We \rev{analyzed all 1200 frames in each of the 1543 available scans using lag-one contrasts in 56-frame (40.32-second) windows}, and performed Hodge decomposition at every \rev{0.72}-second interval (\rev{1 TR}). The magnitude of an edge flow \(X\) was quantified by its \(\ell_2\)-norm, \(\|X\|\). Since the gradient and harmonic flows are mutually orthogonal, we have
\[
\rev{\| X \|^2 = \| X_G \|^2 + \| X_H \|^2.}
\]

\rev{To compare repeat-scan patterns independently of component magnitude, we normalize each nonzero component within each scan and window:}
\[
\rev{X_G(t)\leftarrow\frac{X_G(t)}{\|X_G(t)\|},
\qquad
X_H(t)\leftarrow\frac{X_H(t)}{\|X_H(t)\|}.}
\]

\rev{Normalization preserves relative edge weights and interaction directions, retaining the cyclic pattern while making it invariant to uniform positive rescaling. We average these unit vectors across each subject's available repeat scans at the same window position, denoted by bars, and compute}
\[
\rev{\eta_G(t)=\frac{\|\bar X_G(t)\|^2}
{\|\bar X_G(t)\|^2+\|\bar X_H(t)\|^2},
\qquad \eta_H(t)=1-\eta_G(t).}
\]
Here, \rev{\(\eta_G\) and} \(\eta_H\) measure the relative contributions of the gradient and harmonic components, respectively, and satisfy \rev{\(\eta_G + \eta_H = 1\) \cite{haruna.2016}.
For 399 subjects with repeat scans} (Fig.~\ref{Fig:loopanalysis})\rev{, least-squares fits of $y(t)=\alpha+\beta t+\varepsilon(t)$ at 21 nonoverlapping 56-frame windows (open circles; 56-TR step) showed a $72.5\%$ reduction in slope magnitude after normalization: $-0.921$ to $-0.253$ percentage points per minute for harmonic and $+0.921$ to $+0.253$ for gradient ratios.

We tested $H_0:\beta=0$ by likelihood ratio with estimated AR(1) covariance \cite{beach.1978}. Under joint spectral sign symmetry \cite{winkler.2014}, we fixed magnitudes and refitted both models for all $2^{20}$ coefficient-sign configurations in the centered covariance eigenbasis, avoiding Gaussianity. The resulting $p=113852/2^{20}=0.1086$ indicated no significant linear trend. Normalized trajectories were similar with and without overlap, consistent with lagged-correlation identification} (Fig.~\rev{\ref{fig:cyclic-identification-sensitivity}).

Across subjects and all 1145 windows, the mean normalized energy ratios were \(53.35 \pm 6.04\%\) for the gradient and \(46.65 \pm 6.04\%\) for the harmonic component.}

\begin{table}[t!]
\centering
\caption{Top \rev{10} dominant harmonic flows across brain regions. \rev{Edges are ranked by absolute population mean weight after normalization within each window; arrows follow the sign of the mean interaction contrast}.}
\label{table:20harmonic}
\scriptsize
\begin{tabular}{r l c l c}
\hline
\rev{Rank} & \rev{From} &  & \rev{To} & \rev{Magnitude} \\
\hline
\rev{1} & \rev{Cerebelum-Crus1-L} & \rev{$\to$} & \rev{Vermis-10} & \rev{0.0066} \\
\rev{2} & \rev{Cerebelum-Crus2-L} & \rev{$\to$} & \rev{Amygdala-R} & \rev{0.0064} \\
\rev{3} & \rev{Cerebelum-Crus1-R} & \rev{$\to$} & \rev{Vermis-10} & \rev{0.0058} \\
\rev{4} & \rev{Cerebelum-Crus2-L} & \rev{$\to$} & \rev{Amygdala-L} & \rev{0.0058} \\
\rev{5} & \rev{Cerebelum-Crus1-L} & \rev{$\to$} & \rev{Rectus-L} & \rev{0.0057} \\
\rev{6} & \rev{Cerebelum-Crus1-L} & \rev{$\to$} & \rev{Amygdala-R} & \rev{0.0056} \\
\rev{7} & \rev{Cerebelum-Crus1-R} & \rev{$\to$} & \rev{Amygdala-R} & \rev{0.0055} \\
\rev{8} & \rev{Cingulum-Mid-R} & \rev{$\to$} & \rev{Cerebelum-Crus1-L} & \rev{0.0054} \\
\rev{9} & \rev{Cerebelum-Crus1-R} & \rev{$\to$} & \rev{Amygdala-L} & \rev{0.0053} \\
\rev{10} & \rev{Precuneus-R} & \rev{$\to$} & \rev{Cerebelum-Crus2-L} & \rev{0.0053} \\
\hline
\end{tabular}
\end{table}

\subsection{\rev{Population-Level Cyclic Interactions} in rs-fMRI}
\label{sec:population-cyclic-results}

\begin{figure*}[t]
\centering
\includegraphics[width=\textwidth]{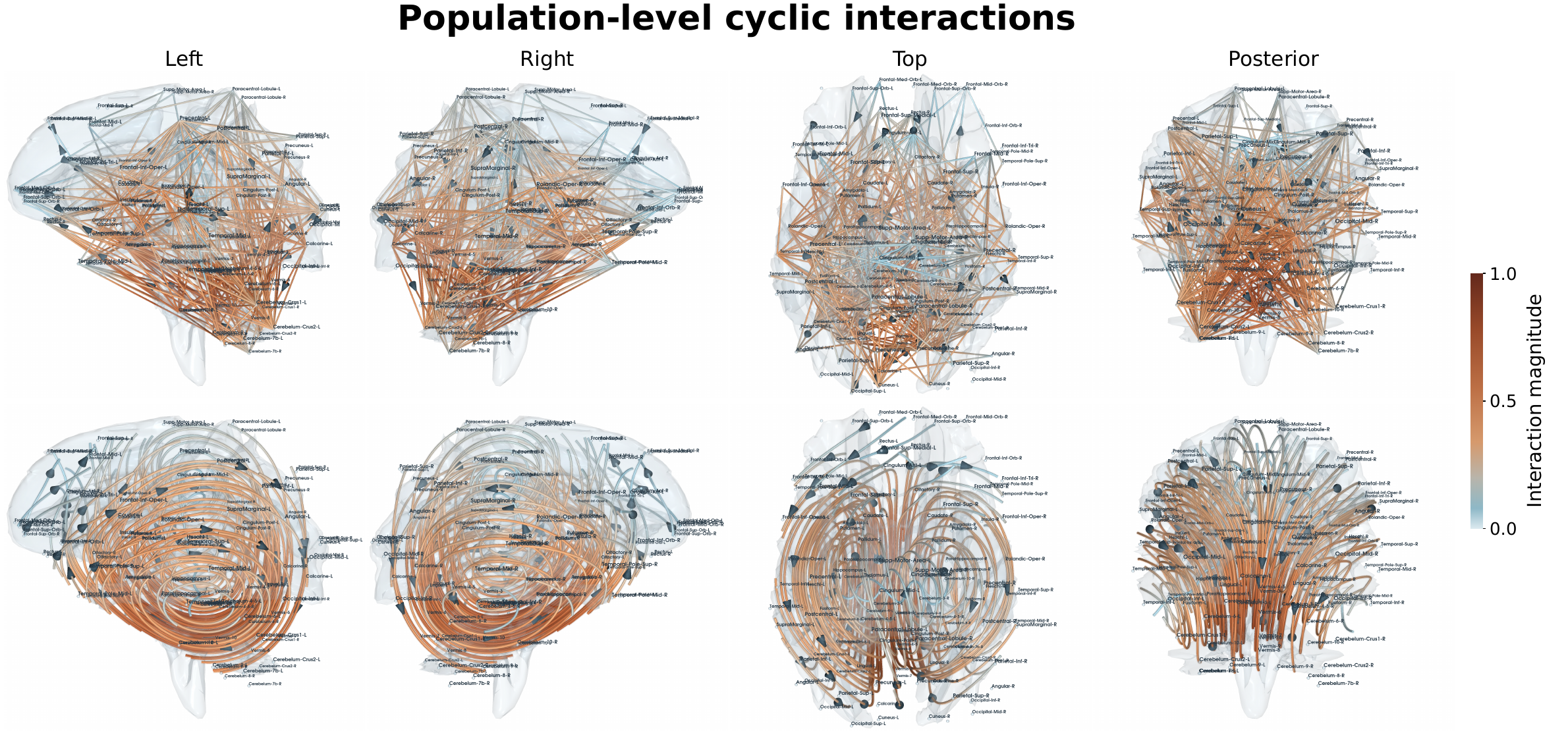}
\caption{{\bf \rev{Population-level cyclic interactions}}\rev{: lag-one harmonic interactions were normalized to unit length within each window, averaged over windows and available scans within each subject, and then averaged across 400 subjects with equal weights (1543 scans). Each window contained 56 frames (40.32 seconds) and advanced by one TR.} {\bf \rev{Top}}\rev{: the 25 directed cycles with the largest coefficients in the cycle decomposition.} {\bf \rev{Bottom}}\rev{: streamlines computed from the full population cyclic interaction field, showing broadly symmetric bilateral circulation and trajectories extending between frontal and posterior regions.}}
\label{fig:population-cyclic-interactions}
\end{figure*}

Harmonic \rev{components---representing} equilibrium states of \rev{the proposed diffusion---capture} dominant cyclic patterns\rev{. The energy--ratio analysis showed a mean normalized harmonic ratio of \(46.65\%\) across} subjects and time, \rev{reflecting repeat-scan pattern consistency}. To visualize the cohort-level structure of these cycles, we averaged the \rev{unit-normalized} harmonic edge flows \(X_H(t)\) over time \rev{within each scan, then across each subject's available repeat scans, and finally across the 400 subjects with equal weights}, and displayed the resulting mean harmonic flow (Fig.~\rev{\ref{fig:population-cyclic-interactions}).

We assessed the statistical reliability of the population cyclic pattern using the one-sample test (Section~\ref{sec:population-spectral-test}) on the subject-level means of these normalized lag-one harmonic interactions from the 400 subjects. We evaluated $B=100{,}000{,}000$ independent sign configurations. No randomized statistic reached the observed value, giving the $p$-value of $1/(B+1)\approx10^{-8}$ and supporting a shared signed population cyclic pattern.}

A large positive harmonic value on edge \(i\!\to\!j\) indicates that the edge participates in \rev{balanced circulation in the population mean. The direction of each of the 10 strongest interactions listed in Table~\ref{table:20harmonic} is consistent across the four scan-specific population averages.} Several strong harmonic flows involve the vermis and \rev{cerebellar Crus~I and II}, together with limbic nodes such as the rectus \rev{and amygdala}. These connections align with \rev{cerebello--limbic loops described in resting-state functional connectivity studies \cite{habas.2021,sang.2012}.}

\rev{Table~\ref{tab:hemi_distribution}} quantifies the distribution of \rev{population harmonic interaction magnitude} across hemispheric and midline systems. \rev{Percentages are calculated by summing absolute population mean weights in each category and dividing by their total across all 6670 edges}. Cross-hemispheric flows (Left $\to$ Right and Right $\to$ Left) together account for \rev{43.60\%. Midline structures---the vermis regions---account for approximately 11.90\% of interaction magnitude.} Overall, this distribution highlights that cyclic interactions are not confined to a single hemisphere but span left, right, and midline systems.

\begin{table}[t!]
\centering
\caption{\rev{Distribution of population harmonic interaction magnitude (\%). Rows indicate source groups; columns indicate target groups. Midline denotes eight cerebellar vermis regions (I--II, III, IV--V, VI--X).}}
\label{tab:hemi_distribution}
\small
{\color{revisionorange}
\begin{tabular*}{\linewidth}{@{\extracolsep{\fill}}lrrr@{}}
\hline
Source / target & Left & Right & Midline \\
\hline
Left    & 22.43 & 21.83 & 2.87 \\
Right   & 21.77 & 22.07 & 2.89 \\
Midline &  2.92 &  2.84 & 0.38 \\
\hline
\end{tabular*}}
\end{table}

\subsection{\rev{Reproducibility and Sensitivity of Cyclic Patterns}}
\label{sec:cyclic-reproducibility}

\begin{figure}[t]
\centering
\includegraphics[width=\linewidth]{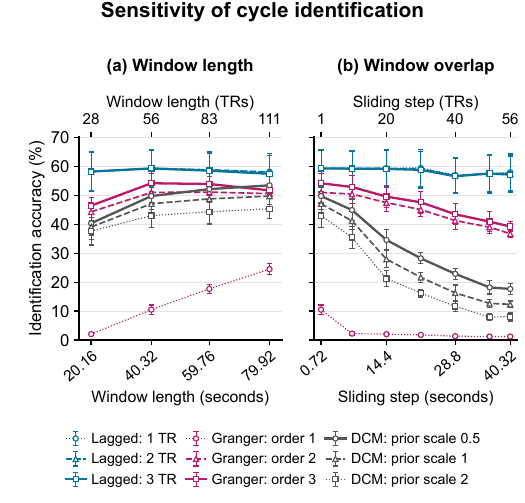}
\caption{{\bf \rev{Sensitivity of cycle identification}}\rev{: mean identification accuracy and sample SD across six repeat-scan pairs using the regionally weighted cosine-based score in Eq.~\ref{eq:regional-cyclic-score}, with weights fitted only to training subjects.
(a) Window-length sensitivity with one-TR sliding. (b) Overlap sensitivity with a fixed 56-frame (40.32-second) window; increasing the sliding step reduces overlap.} }
\label{fig:cyclic-identification-sensitivity}
\end{figure}

\rev{We jointly assessed empirical reproducibility and sensitivity using four repeat HCP rs-fMRI scans from the original 400-subject cohort, giving $\binom{4}{2}=6$ scan-pair comparisons. Available subjects were retained for each pair. We estimated directed interactions using lagged correlation \cite{mitra.2014}, conditional Granger causality as described in Section~\ref{sec:estimation-recovery} \cite{barnett.2014,friston.2014,seth.2015}, and DCM \cite{frassle.2021,friston.2013}, and applied the same harmonic projection to their forward--backward differences. Harmonic interactions were averaged over sliding windows.

For each scan pair, we tested whether a subject's harmonic pattern identified the same subject in the other scan. This task is motivated by connectome fingerprinting \cite{finn.2015}. We used five-fold cross-validation, keeping all scans from each subject together. Depending on scan availability, each fold contained 68--86 eligible test subjects.

For scan $s$, let $\bar X_{H,s}$ denote the temporal mean of $X_H(t)$. After subtracting its mean across edges, we normalize $\bar X_{H,s}\leftarrow\bar X_{H,s}/\|\bar X_{H,s}\|$. The regionally weighted cosine-based score between scans $s$ and $t$ is}
\begin{equation}
\rev{\label{eq:regional-cyclic-score}
\rho(s,t)=\bar X_{H,s}^\top W_C\bar X_{H,t},}
\end{equation}
\rev{where $W_C$ is diagonal, with each edge weighted by the average of the learned weights of its two endpoint regions. Setting all weights to one gives ordinary cosine similarity between the centered vectors. Cosine similarity compares cyclic patterns independently of their overall magnitude \cite{chung.2026.EMBC.causality,luo.2018}. Because regions differ in how well their connectivity patterns identify individuals \cite{finn.2015}, we learn regional weights to emphasize interactions that distinguish subjects across repeat scans. These weights are shared across subjects and scans within each fitted model.

For a scan pair with $n$ training subjects who have both scans available, let $s_i$ and $t_i$ denote the two scans of subject $i$, and write $\rho_{ij}=\rho(s_i,t_j)$. Diagonal entries compare the same subject; off-diagonal entries compare different subjects. We learn the regional weights by minimizing the bidirectional cross-entropy \cite{chen.2020} with a penalty toward equal weighting:}
\begin{equation}
\rev{\label{eq:cyclic-identification-objective}
\ell(W_C)+\lambda\|W_C-I\|_F^2,}
\end{equation}
\rev{where}
$$
\rev{\begin{aligned}
\ell(W_C)=-\frac{1}{2n}\sum_{i=1}^{n}\Bigg[
\log\frac{e^{\rho_{ii}/\eta}}{\sum_{j=1}^{n}e^{\rho_{ij}/\eta}} + \log\frac{e^{\rho_{ii}/\eta}}{\sum_{j=1}^{n}e^{\rho_{ji}/\eta}}\Bigg].
\end{aligned}}
$$
\rev{Each fraction assigns a probability to the correct identity among $n$ candidates, and the two terms evaluate matching in opposite scan directions. Cyclic interactions may vary across time and subjects while retaining similar overall patterns. A high cosine similarity alone therefore fail to distinguish a subject's repeated scans from those of other subjects. Cross-entropy penalizes high scores for incorrect matches \cite{chen.2020}, guiding the regional weights toward reproducible features that distinguish each subject. The temperature $\eta$ controls the sharpness of the matching probabilities, and $\lambda$ controls the penalty toward equal weighting.

For each interaction model and setting, we used an adaptive tree search over  $\eta\in[10^{-4},1]$ and $\lambda\in[10^{-12},1]$ \cite{munos.2011}. We selected the parameter pair that maximized identification accuracy. All scans from held-out subjects were excluded from these steps. Each held-out subject was matched to the highest-scoring eligible held-out candidate in the other scan. Identification accuracy was calculated for each pair by pooling correct matches across folds and both scan directions. Figure~\ref{fig:cyclic-identification-sensitivity} reports the mean accuracy and sample SD across the six pairs.}

\rev{We examined} {\it \rev{sensitivity to window length and overlap}} \rev{because these choices might affect the subject-level harmonic patterns used for identification. Window length determines the trade-off between temporal resolution and the observations available for estimation \cite{leonardi.2015}. Recording duration also affects the reliability of resting-state connectivity estimates \cite{birn.2013,ma.2024}. We therefore tested window lengths of 20, 40, 60, and 80 seconds with one-TR sliding. To assess dependence on densely overlapping windows, we separately increased the sliding step from one TR to nonoverlapping windows while fixing the window at 56 frames (40.32 seconds). In both analyses, we also varied correlation lags and Granger orders over 1, 2, and 3 to examine sensitivity to the temporal offset and modeled history, respectively. For DCM, we varied coupling-prior variance scales over 0.5, 1, and 2 to assess sensitivity to stronger or weaker regularization of the estimated interactions.

Lagged correlation showed relatively stable identification accuracy across all tested settings. Granger and DCM were more sensitive to reduced overlap, which changes both the number and placement of windows contributing to each scan average. At the nominal 40-second window with one-TR sliding, harmonic patterns derived from lag-one correlation achieved 59.25\% identification accuracy, exceeding Granger order 3 and DCM prior scale 0.5 by 5.13 and 9.56 percentage points, respectively.}

\section{Conclusion \& Discussion}

\subsection{\rev{Cyclic Interactions vs. Higher-Order Interactions}}
\rev{The proposed harmonic flow} is only possible when the network has nontrivial \emph{first cohomology} indicating it contains \rev{closed} paths that cannot be collapsed to a point. The harmonic subspace is \rev{$\mathcal{H}^1=\ker(\mathbf B_1)=\ker(\mathcal L_1)$}, whose dimension is the first Betti number $\beta_1$, which \rev{counts} the number of independent noncontractible cycles \rev{\cite{lim.2020}}. In a simply connected graph ($\beta_1=0$) there is no harmonic flow\rev{.} Networks that exhibit clear “ring” pathways around holes \rev{support harmonic circulation. Cyclic interactions are one-dimensional structures formed by closed paths of directed edges; an arbitrary collection of directed edges does not necessarily form such paths. Estimating individual node-to-node interactions does not by itself isolate this balanced cyclic component \cite{anand.2024.ISBI}. We extract it from directed networks through the minimum-energy principle \cite{chung.2026.ISBI}.

Lagged correlation estimates each interaction from a pair of time series, whereas conditional Granger models histories from all measured nodes. These additional predictors account for conditional dependencies but increase the number of parameters to estimate without explicitly isolating cyclic structure. In our simulations, lagged correlation achieved better cyclic-pattern recovery than conditional Granger (Section~\ref{sec:estimation-recovery}). It also achieved higher repeat-scan identification accuracy (Section~\ref{sec:cyclic-reproducibility}). Thus, conditioning on all measured nodes does not by itself guarantee better cyclic-pattern recovery or reproducibility.

Existing information-theoretic approaches already characterize multivariate and higher-order dependencies \cite{antonacci.2021,antonacci.2025,antonacci.2025.JNE,faes.2022,sparacino.2025,pirovano.2023,faes.2025}. An edgewise estimate can already contain higher-order information. For example, partial information decomposition (PID) expresses transfer entropy as information unique to the source region's history plus synergy between the source and target histories. Investigating the relationship between these information theoretic approaches and harmonic circulation is left for future study.}

\subsection{Biological Interpretation of Harmonic Flow}

\rev{Harmonic flow is} the steady-state component of \rev{the proposed diffusion dynamics: gradient interactions associated with source--sink imbalance dissipate, while balanced circulation remains unchanged \cite{chung.2026.ISBI}. This distinction motivates examining whether the preserved cyclic patterns capture reproducible organization in} resting-state \rev{brain activity}.

\rev{Our rs-fMRI results support this interpretation through repeat-scan reproducibility and population-level consistency. The extracted patterns also exhibit two anatomically meaningful features}. First, \rev{cross-hemispheric interactions account for $43.60\%$ of the population interaction magnitude, indicating that cyclic organization extends across hemispheres. Interhemispheric coordination is also evident in resting-state functional connectivity \cite{stark.2008,jin.2020}. Second, strong interactions involve the vermis, cerebellum, and limbic regions, linking midline and bilateral structures within these patterns}. Together, these \rev{findings} suggest that harmonic flow \rev{captures a reproducible component of distributed} resting-state connectivity.

\subsection{\rev{Limitations of the Study}}
\label{sec:limitations}
\label{sec:relation-causal-frameworks}

\rev{Our framework provides a dynamical and energetic description of cyclic organization in spontaneous resting-state brain activity, without claiming intervention effects in Pearl's framework \cite{pearl.2009.overview}. Directed interactions are estimated before harmonic projection, which retains each estimator's interpretation: temporal association for lagged correlation, predictive dependence for conditional Granger causality \cite{seth.2015}, and effective connectivity under a generative model for DCM \cite{friston.2013}. The energy model describes relaxation toward balanced circulation; it neither equates these interpretations nor supplies causal identification assumptions. Repeat-scan identification and population-level inference support reproducibility and consistency, without establishing neural feedback circuits.

Averaging voxel signals within each of the 116 AAL regions can obscure distinct temporal patterns within a region. Consequently, cyclic interactions are resolved only between these regions, and their localization and reproducibility at finer parcellations have not been established.

We identified the cerebellar--limbic and cross-hemispheric patterns without incorporating anatomical connectivity. Their correspondence to white-matter pathways therefore remains to be tested.} When structural connectivity from diffusion-MRI is available, the Dirichlet energy can be modified to incorporate anisotropy in white-matter tracts. Edges are assigned conductance $C=(c_{ij})$ reflecting the strength of axonal connectivity along that tract. The tract conductance can be estimated from streamline counts, mean fractional anisotropy (FA), or related microstructural measures \rev{\cite{chung.2017.BC,honey.2009,gong.2009}. Tract conductances could constrain a weighted Hodge model. The present model minimizes interaction imbalance, without optimizing wiring length or communication efficiency. Such an extension would connect interaction organization to the trade-off between wiring cost and communication capacity \cite{bullmore.2012}, including the role of long-range connections in shortening communication paths \cite{kaiser.2006}.} We leave the systematic exploration of this structure-informed extension as a future study\rev{. Our controlled simulations assess recovery against known cyclic ground truth; evaluation on social or communication networks with established benchmarks remains future work}.

\makeatletter
\begingroup
\let\originalbibitem\bibitem
\def\addedreferences{abdelnour.2014,antonacci.2021,antonacci.2025,antonacci.2025.JNE,barnett.2014,beach.1978,bullmore.2012,chen.2020,chung.2017.BC,chung.2019.transposition,chung.2026.EMBC.causality,chung.2026.ISBI,faes.2022,faes.2025,finn.2015,friston.2010,friston.2013,hemerik.2018,hopfield.1984,kaiser.2006,leonardi.2015,mitra.2014,munos.2011,parikh.2014,pearl.2009.overview,pirovano.2023,schaefer.2018,seth.2015,sparacino.2025,wang.2022,winkler.2014}
\renewcommand{\bibitem}[1]{%
  \color{black}\def\currentreference{#1}%
  \@for\addedreference:=\addedreferences\do{%
    \ifx\currentreference\addedreference\color{revisionorange}\fi}%
  \originalbibitem{#1}}
\bibliographystyle{IEEEbib-revision}
\bibliography{revision1-references}
\endgroup
\makeatother
\end{document}